\documentclass[12pt,preprint]{aastex}
\received{}
\revised{}
\accepted{}

\shorttitle{H~1821}
\shortauthors{Fang et al.}

\begin{document}

\title{A {\sl Chandra} HETG Observation of the Quasar H~1821+643 and Its Surrounding Cluster}

\author{Taotao Fang, David S. Davis, Julia C. Lee, Herman L. Marshall, Greg L. Bryan, AND Claude R.Canizares}
\author{}
\affil{Department of Physics and Center for Space Research}
\affil{Massachusetts Institute of Technology}
\affil{NE80-6081, 77 Massachusetts Avenue, Cambridge, MA 02139}
\email{fangt@space.mit.edu}

\begin{abstract}

We present the high-resolution X-ray spectrum of the low-redshift quasar H~1821+643 and its surrounding hot cluster observed with the {\sl Chandra} High Energy Transmission Grating Spectrometer (HETGS). An iron emission line attributed to the quasar at $\sim 6.43$ keV (rest frame) is clearly resolved, with an equivalent width of $\sim 100$ eV. Although we cannot rule out contributions to the line from a putative torus, the {\sc diskline} model provides an acceptable fit to this iron line. We also detect a weak emission feature at $\sim 6.9$ keV (rest frame). We suggest that both lines could originate in an accretion disk comprised of a highly ionized optically thin atmosphere sitting atop a mostly neutral disk. We search for absorption features from a warm/hot component of the intergalactic medium along the $\sim 1.5h^{-1}$ Gpc line of sight to the quasar. No absorption features are detected at or above the $3\sigma$ level while a total of six \ion{O}{6} intervening absorption systems have been detected with {\sl HST} and {\sl FUSE}. Based on the lack of \ion{O}{7} and \ion{O}{8} absorption lines and by assuming collisionally ionization, we constrain the gas temperature of a typical \ion{O}{6} absorber to $10^{5} < T < 10^{6}$ K, which is consistent with the results from hydrodynamic simulations of the intergalactic medium. The zeroth order image reveals the extended emission from the surrounding cluster. We have been able to separate the moderate CCD X-ray spectrum of the surrounding cluster from the central quasar and find that this is a hot cluster with a temperature of $\sim 10$ keV and a metal abundance of $\sim 0.3Z_{\odot}$. We also independently obtain the redshift of the cluster, which is consistent with the optical results. We estimate that the cluster makes negligible contributions to the 6.9 keV iron K line flux.

\end{abstract}


\clearpage

\section{Introduction \label{sec:intro}}

H~1821+643 is one of the most luminous quasars ($m_{v} = 14.1$) at low redshift ($z = 0.297$). It was discovered as a serendipitous X-ray source detected with the {\sl Einstein Observatory} \citep{pma84} and has been studied extensively (e.g., \citealp{khs91, bla95, sbt97}). It has been classified as a radio-quiet quasar according to its radio luminosity and nuclear [\ion{O}{3}] luminosity \citep{lrh92}. H~1821+643 is located in a cluster of galaxies with Abell richness class $\geq 2$ \citep{sbg92}.

Previous observations have shown an iron K$\alpha$ emission line in the X-ray spectrum of H~1821+643. {\sl Ginga} (\citealp{kwo91, khs91}) observed an iron line at $6.6 \pm 0.3$ keV (rest frame of the quasar). The subsequent {\sl ASCA} observation improved the measurement, and gave an observed line center energy of $5.07 \pm 0.04$ keV, or $6.58 \pm 0.05$ keV in the quasar rest frame \citep{ymi97}, and a line width of $\sim 160$ eV. The authors attributed this to a broadened fluorescent iron line from a highly-ionized accretion disk. A particular problem with these observations is that although extened X-ray emission from the cluster that surrounds H~1821+643 has been detected with {\sl ROSAT} (\citealp{sbt97, heg97}), none of these observations has been able to obtain the spectrum of the cluster and therefore separate the contribution to the quasar iron line from the cluster. In a {\sl ROSAT} PSPC observation \citet{sbt97} concluded that the cluster could contribute $10-70\%$ to the observed Fe emission line with {\sl Ginga}. \citet{ymi97} placed an upper limit to the cluster contribution at $<7\%$ by assuming a metal abundance of $< 0.4 Z_{\odot}$ and a thermal temperature of $kT <10$ keV.

In this paper we present a high-resolution X-ray spectrum of H~1821+643 obtained with the {\sl Chandra} High Energy Transimission Grating Spectrometer (HETGS). We clearly resolve an emission line from neutral or low-ionized iron. Using the high spatial resolution of {\sl Chandra} we have been able to resolve the extended emission from the surrounding cluster and for the first time obtain its X-ray spectrum. This enables us to determine the cluster contribution to the central quasar and make a more accurate line diagnostic.

Another motivation of this observation is to detect possible X-ray absorption features introduced by the warm/hot intergalactic medium in the local universe. At a moderate redshift, the sight line toward H~1821+643 traverses a distance of nearly $1.5h^{-1}$ Gpc (we use a Hubble constant of $H_{0} = 100h\ {\rm km\ s^{-1}Mpc^{-1}}$ and $h=0.5$). Numerous absorption systems (such as H, C, N, O, Si, etc.) have been discovered in this sight line in optical and UV bands (e.g., \citealp{tls98}). Especially in recent observations, the {\it Hubble Space Telescope} ({\sl HST}) and the {\it Far Ultraviolet Spectroscopic Explorer} ({\sl FUSE}) have detected a number of \ion{O}{6} absorption lines (see, e.g.,  \citealp{tsa00, ots00}) which are not clearly associated with any galactic system. Numerical simulations (\citealp{cos99a, dco01, cto01, fbr01}) predict that at least some of this may be in the form of a moderately warm/hot ($\sim 10^{5} < T < 10^{7}$ K) component of the intergalactic medium (IGM) and the baryonic gas contained in the \ion{O}{6} lines is about $20-30\%$ of this warm/hot IGM (WHIM). The remaining $70-80\%$ WHIM gas is in the form of hot gas which may be detected by \ion{O}{7} and \ion{O}{8} absorption lines in the X-ray band. H~1821+643 is one of the best candidates to observe for this purpose, given that it is one of the brightest X-ray quasars and has a relatvely simple spectral shape in the soft X-ray band.

This paper is arranged as follows: \S\ref{sec:da} is data analysis, \S\ref{sec:con} describes the continuum spectrum of H~1821+643, \S\ref{sec:emiss} is devoted to emission line diagnostics, in \S\ref{sec:abs} we discuss the X-ray constraints on the \ion{O}{6} absorbers, and study an absorption system at $z\approx 0.1214$. We discuss the emission from the surrounding cluster in  \S\ref{sec:cluster}. The last section is the summary.

\section{Data Analysis \label{sec:da}}

H~1821+643 was observed with the {\sl Chandra} High Energy Transmission Grating Spectrometer (HETGS; \citealp{can01}) on 2001 February 9. The total exposure time was $100$ ksec. The HETGS produced a zeroth order image at the aim-point on the focal plane detector, the ACIS-S array, with higher order spectra dispersed to either side (for ACIS-S, see \citealp{gar01}). The telescope pointing direction was offset $20^{\arcsec}$ along $+$Y in order to move the zeroth order off a detector node boundary. The Science Instrument Module (SIM) was moved toward the read-out row by about $3\ mm$ to get better ACIS energy resolution (for detailed instrument setups, see the {\sl Chandra} Proposers' Observatory Guide, or POG\footnote{See {\sl Chandra} Proposers' Observatory Guild (POG) at http://asc.harvard.edu.}). Figure~\ref{fig-1} displays the zeroth-order image in a $2^{\arcmin} \times 2^{\arcmin}$ region. We have smoothed the image for visual clarity. The zeroth-order count rate within a circle of $10^{\arcsec}$ radius is $0.23 \pm 0.03\ {\rm counts\ s^{-1}}$ (error is quoted at $1\sigma$ level) and its light curve shows no significant variation during the observation. This count rate gives a pileup fraction of $\sim 20\%$ (see POG) and prevents us from analyzing the zeroth-order spectrum of the nucleus. We also label the position of the star K1-16. The central star of the planetary nebula K1-16 was detected with the {\sl ROSAT} PSPC at about $1.^{\arcmin}3$ north and $0.^{\arcmin}5$ west of H~1821+643 \citep{khs91}. Although this star would only significantly contribute to the ultrasoft emission between $0.07$ and $0.17$ keV, a roll angle of $46^{\circ}$ was applied to keep the spectra of the two sources from overlapping.

\placefigure{fig-1}

Spectral extractions and reductions were performed with the standard pipeline for the {\sl Chandra} HETGS provided by the Chandra X-ray Center (CXC)\footnote{See http://asc.harvard.edu/.}. We used a combination of {\sl Chandra} Interactive Analysis of Observations (CIAO) V2.1 and custom routines in IDL. The standard screening criteria were applied to the data. We selected photon events with {\sl ASCA} grades 0, 2, 3, 4, 6 and excluded those with energies above $10$ keV. The HETGS consists of two different grating assemblies, the High Energy Grating (HEG) and the Medium Energy Grating (MEG), and provides nearly constant spectral resolution ($\Delta\lambda = 0.012 \AA$ for HEG and $\Delta\lambda = 0.023 \AA$ for MEG) through the entire bandpass (HEG: 0.8-10 keV, MEG: 0.4-8 keV). Extraction windows of $10^{\arcsec}.5$ and $7^{\arcsec}.8$ widths in the cross-dispersion direction were applied for HEG and MEG, respectively. The moderate energy resolution of the ACIS-S is used to separate the overlapping orders of the dispersed spectrum. We added the plus and minus sides to obtain the first order spectra of both grating assemblies (see Figure~\ref{fig-2}). The effective area (ARF) for this observation was obtained using the CIAO tools.

\placefigure{fig-2}

\section{Continuum \label{sec:con}}

The spectrum in Figure~\ref{fig-2} shows a smooth continuum with an emission feature at $\sim 5$ keV. In general, we find that a single power law with photoelectric absorption fits the entire bandpass of both MEG and HEG spectra very well. The MEG and HEG spectra were binned to a constant resolution of $0.02\AA$. We initially performed a simultaneous fit for HEG (0.9---8 keV) and MEG (0.6---8 keV) with a single power law plus photoelectric absorption at $z=0$. We ignored the MEG spectrum below 0.6 keV due to $\sim 30-50\%$ systematic uncertainties of spectral fluxes at the oxygen K edge region. The fitting parameters are shown in Table~\ref{tbl-1}. In two fits, we fixed the absorption column density at the Galactic value ($N^{G}_{H} = 3.8 \times 10^{20}\ {\rm cm^{-2}}$; \citealp{lsa95}). Figure~\ref{fig-3} shows the best fit spectra and the ratio (data divided by model). There is some weak evidence for excess emission below 1 keV. When we allow the absorption column density to vary, the best-fit column density $N_{H}$ is lower than $1.9 \times 10^{19}\ {\rm cm^{-2}}$ at $90\%$ confidence level, so we infer that there is excess emission in the soft X-ray band. The free-$N_{H}$ model is a better fit to the data because the C-statistic drops significantly ($\Delta C = 82$) while the number of the degree of freedom drops imperceptibily. However, the soft excess might also result from residual calibration uncertainties in the MEG effective area ($\sim 30 \%$ for $0.5 < E < 0.8$ keV).  

\placetable{tbl-1}
\placefigure{fig-3}

Since the systematic uncertainties of the HETGS spectral fluxes are lower above 2 keV (less than $10 \%$) and an accurate measurement of the continuum level is important in identifying the emission feature, we fit instead the HEG and MEG spectra between 2 and 7 keV. Figure~\ref{fig-2} shows that a single power law fits both MEG and HEG spectra very well in this band. Since this energy band is largely insensitive to $N_{H}$, we fix the column density at the Galactic value. The best-fit photon index ($\Gamma = 1.761^{+0.047}_{-0.052}$) \footnote{Unless mentioned, all the errors are quoted at $90\%$ confidence level.} is consistent with the previous result ($\Gamma = 1.75\pm 0.03$) reported by {\sl ASCA} \citep{ymi97} at better than the $1\sigma$ level. It is also consistent with indices of other radio-quiet quasars \citep{rtu00}. The flux (2---10 keV, observer frame) and luminosity (2---10 keV, quasar frame) are $\sim 1.2 \times 10^{-11}\ {\rm ergs\ cm^{-2}s^{-1}}$ and $\sim 5 \times 10^{45}\ {\rm ergs\ s^{-1}}$, respectively. These values are also consistent with those reported from {\sl ASCA} \citep{ymi97}, {\sl EXOSAT} \citep{wby89}, {\sl Ginga} (\citealp{kwo91, khs91}) and {\sl BBXRT} \citep{ysm93}. {\sl ASCA} observation reported a flux of $\sim 1.75 \times 10^{-11}\ {\rm ergs\ cm^{-2}s^{-1}}$ and a luminosity of $\sim 7 \times 10^{45}\ {\rm ergs\ s^{-1}}$ in the corresponding energy bands.

\section{Emission Lines \label{sec:emiss}}

\subsection{Spectral Fitting \label{sec:spec}}

Figure~\ref{fig-3} shows that the Emission feature at $\sim$ 5 keV can be well fitted with the Gaussian parameters detailed in Table~\ref{tbl-2}. We also fit the HEG and MEG spectra separately to check for consistency, and we list the results from fitting the HEG data only in Table~\ref{tbl-2}. Given that the respective HEG and MEG resolutions are $\sim 40$ eV and 80 eV FWHM (full width at half-maximum) at the observed line energy (POG), the line is clearly resolved by both grating assemblies. Table~\ref{tbl-2} shows that the Gaussian fit significantly improves the C-statistic by $\Delta C = 40.2$, indicating a detection significance of nearly $6\sigma$ for three degrees of freedom (namely, Gaussian line flux, line width and the center energy). To investigate whether or not the line is the neutral K$\alpha$ line, we also fit the spectra by fixing the line center energy at 6.4 keV. The C-statistic is higher by 1.28 with one less degree of freedom, which means a neutral iron K$\alpha$ line can be ruled out at $68\%$ level but not at $90\%$ level. Figure~\ref{fig-4} shows the best fit and $68\%$, $90\%$ and $99\%$ joint confidence regions for the line center energy and the line flux. 

\placetable{tbl-2}
\placefigure{fig-4}

The iron K$\alpha$ line feature in H~1821+643 was measured with {\sl Ginga} observations \citep{kwo91} at $6.6 \pm 0.3$ keV (rest frame of the quasar) and with {\sl ASCA} at $6.58 \pm 0.05$ keV \citep{ymi97}. {\sl Chandra} observations give a rest-frame line energy of $6.435 \pm 0.041$ keV ($6.435^{+0.197}_{-0.106}$ keV at $3\sigma$ level). This shows that the line energy from the {\sl ASCA} observation was excluded at $90\%$ confidence but is within $3\sigma$ of the {\sl Chandra} observation. 

Even after the addition of this iron K$\alpha$ line, there are still residuals present at energies $> 6.4$ keV (quasar frame) in both HEG and MEG spectra. We add a second Gaussian line component, with the energy and the line width as free parameters. The spectral fit results for this second component are also listed in Table~\ref{tbl-2}. The C-statistic decreases by $\Delta C = 9.7$, which corresponds to a detection significance of $2.3\sigma$. Fitting with the HEG data only fails to constrain the line width; however, the simultaneous spectral fit of HEG and MEG gives a line width of $\sim 61$ eV.  

We have used the extracted cluster profile and spectrum to estimate the amount of flux that the cluster could have contributed to both emission lines in the grating spectra (see \S\ref{sec:cluster}). We conclude that the hot cluster should have no contribution to the neutral iron line at $\sim 6.4$ keV (rest frame). Using the profile derived in \S\ref{sec:cluster} we find that the cluster can contribute at most $3\%$ flux to the 6.9 keV line.

\subsection{The 6.4 KeV Line \label{sec:6.4}}

The observed line energy of $\sim$ $6.4$ keV implies that the fluorescent iron line is from neutral and/or low ionization states of iron, although the $90\%$ confidence errors cannot rule out contribution from ionization states up to \ion{Fe}{19} (see, e.g., \citealp{mgv85}). The line center energy is consistent with the energy of \ion{Fe}{3} K$\alpha_{1}$ (6.434 keV) or \ion{Fe}{4} K$\alpha_{2}$ (6.435 keV) (see, e.g., \citealp{kme93}); however, we cannot rule out \ion{Fe}{1} K$\alpha$ emission from an outflow with a velocity of $v \sim 2000\ {\rm km\ s^{-1}}$. 

Such a fluorescent iron line implies the existence of cold ($T<10^{6}$ K) reprocessing material (see, e.g., \citealp{fir00}) illuminated by X-ray continuum. An incident X-ray photon is either Compton scattered by electrons in the cold gas, or photoelectrically absorbed followed by fluorescent line emission or Auger de-excitation. This photoelectric absorption would produce iron K edges at relatively higher energies. Iron K edges were not detected in the HEG and MEG spectra of H~1821+643. In order to estimate the significance of the iron K edge, we include a multiplicative edge model to the power law plus the Gaussian model. The edge energy is fixed at 7.1 keV (quasar frame) because the K edges of most neutral and low ionization states of iron (up to \ion{Fe}{5}) are around this energy \citep{vya95}. The $90\%$ upper limit to the optical depth of the edge is $\tau \sim 6.2\times 10^{-2}$, which corresponds to a $90\%$ upper limit of the hydrogen equivalent column density at $\sim 6 \times 10^{22}\ {\rm cm^{-2}}$. Here we assume solar abundance \citep{agr89} and use the photoionization cross sections from \citet{vya95}.             

Several models have been proposed to explain the origin of the reprocessing material responsible for the Fe line. Three possible scenarios are that the iron line is reflected (1) from material in the disk close to the black hole which is subject to relativistic and transverse Doppler effects ({\sc diskline} model, \citealp{frs89}), (2) from a molecular torus \citep{ant93}, (3) or from the broad line region (BLR) clouds \citep{ygn01}. We find that the {\sc diskline} model gives an acceptable fit to the iron emission line. However, we cannot rule out the torus model if the reflection material has a small Thomson optical depth. A BLR origin for the Fe line requires a strong flux variation on timescales of $\sim$ months. Here we discuss each model.

{\bf {\sc diskline} model:} The {\sc diskline} model provides acceptable fits to both HEG and MEG data. Thus far, this has been the most commonly used model for explaining the broad Fe K$\alpha$ line observed in many Seyfert 1 galaxies \citep{ngm97} and quasars \citep{rtu00}. In this model, X-rays are reprocessed in a cold or highly ionized accretion disk rotating around a Schwarzschild black hole with a characteristic radius of $r_{g} = GM/c^{2}$ (see, e.g., \citealp{frs89}), where $M$ is the mass of the central black hole. The intrinsically narrow line (line width $< 1$ eV) is broadened by the relativistic motion of the disk ranging from a few hundred eV to $\sim$ keV (see \citealp{fir00} for a review). We fix the inner and outer disk radii at $6r_{g}$ and $1000r_{g}$ respectively, where $6r_{g}$ is the last stable orbit radius of a Schwarzschild black hole with mass $M$. The radial power law index of the emissivity is fixed at $\alpha=-2$. Three parameters, the line centroid energy, the line intensity and the disk inclination angle ($\theta$), are allowed to vary. The spectral fit results are shown in Table~\ref{tbl-2}. The C-statistic of the {\sc diskline} model is comparable to that of a simple Gaussian fit, implying both models are acceptable. The line centroid energy is $\sim 6.49$ keV, comparable to within the $90\%$ confidence error of the best-fit Gaussian value reported earlier. The {\sc diskline} model predicts a skewed, double-horned line profile in which the blueward side drops sharply. However, our {\sl Chandra} observation is not able to detect this level of structure.

Iron K emission line seen with {\sl ASCA} shows an EW of $170\pm50$ eV at 6.58 keV \citep{ymi97}. One of their explanations for the line strength and position is that the line is reprocessed from an accretion disk in which iron is ionized up to the He-like states ($\sim 6.68$ keV) in the innermost part, and redshifted to the observed 6.58 keV energy by the strong gravitation of the central black hole. We suspect that the {\sl ASCA} data was unable to distinguish between the 6.4 and 6.9 keV lines, given its resolution (the {\sl ASCA} resolution at the Fe K energies is $\sim 160$ eV FWHM, respectively 4 and 2 times worse when compared with the HEG and MEG). The {\sl Chandra} data clearly show an emission line from neutral or low ionization states.

{\bf Torus:} Recent {\sl Chandra} HETGS and {\sl XMM-Newton} observations have shown that narrow iron K emission at $\sim 6.4$ keV is prevalent in many active galactic nuclei (AGNs) and quasars (e.g., \citealp{rtp01} for Mkn 205, \citealp{pro01} for Mkn 509 and \citealp{ygn01} for NGC 5548). A common phenomenon is that these AGNs and quasars showed a broader Fe K emission in their {\sl ASCA} spectra. For Mkn 205 and Mkn 509 the narrow component is attributed to reflection from the putative torus based on the $\sim 50-80$ eV measured EWs. According to \citet{kmz94}, the the characteristic EW of an iron K$\alpha$ line from the torus could be $\sim 100$ eV if the reflection material has a typical Thomson optical depth $\tau_{T} = 0.5 - 1$. The EW in H~1821+643 is $\sim 110$ eV.

{\bf BLR clouds:}  The line width of the 6.4 keV line in H~1821+643 is $\sigma = 106.9^{+58.3}_{-37.8}$ eV, which corresponds to $1.171^{+0.639}_{-0.414} \times 10^{4}\ {\rm km\ s^{-1}}$ FWHM. \citet{ygn01} observed a narrower line ($\sigma \approx 4525\ {\rm km\ s^{-1}}$ FWHM) in NGC 5548 with the {\sl Chandra} HETG. They attributed the broadening of the Fe K$\alpha$ line to the bulk velocity of the emitting gas, which may be located in the broad line region. Based on their model, we calculate the line intensity and EW of the 6.4 keV line in H~1821+643, using the upper limit of the hydrogen column density from the iron K-edge constraint. We obtain the $90\%$ upper limit of the line intensity and EW of $\sim 5\times 10^{-6}\ {\rm photons\ cm^{-2}s^{-1}}$ and $\sim 25$ eV, respectively. Both values fall short of the observed values by a significant factor of $\sim 5$. \citet{ygn01} encountered the same problem and they attributed this to the possible variation of the continuum level shortly before the {\sl Chandra} observation. For H~1821+643, on a central black hole mass of $\sim 10^{9}\ M_{\odot}$ \citep{khs91}, the BLR regions which emit the iron K line should be located at a distance of $50.1^{+69.7}_{-29.1}$ light days from the central X-ray source. In order to reconcile the observed EW with that expected from the BLR, the X-ray flux of H~1821+643 would have to have varied by a factor of $\sim 5$ in a timescale of $\sim 2$ months. {\sl ASCA} and {\sl Chandra} observations showed that H~1821+643 experienced only a $\sim 30\%$ flux variation in a timescale of a few years. So we consider a fivefold variation to be unlikely and conclude that the emission from BLR is not the main source of the observed iron K line.

\subsection{The 6.9 keV Line \label{sec:6.9}}

There are several possibilities for the origin  of the \ion{Fe}{26} emission line detected  at 6.9~keV. These include environments (e.g. narrow line regions, or hot accretion disk) which are hot enough,  or have high enough ionization to  produce H-like Fe. However, it is of interest to consider a scenario which can accommodate both the  Fe~K$\alpha$ and \ion{Fe}{26} lines seen in the  Chandra HETGS spectrum for this source, since this could provide us with the unique prospect for probing in detail the condition and structure of the accretion disk in H~1821+643.

Irradiation by X-rays can photoionize the surface layers of the accretion disk (e.g. \citealp{rfa93, rfy99}). The iron line(s) produced by the illuminated matter and the associated reflection spectrum depend largely on the ionization parameter $\xi$ (e.g. \citealp{mfr93, mfr96}, for the case of a constant density structure accretion disk). There are also other dependences, e.g. the spectral index of the incident X-rays (the maximum gas temperature is indirectly related to $\Gamma$), and the strength of the illumination (characterized by the gravity parameter $A$, respectively for high and low illumination, $A \lesssim 0.1$, and $A \gg 1$). Recent calculations of ionized reflection spectra which include hydrostatic equilibrium effects  have been presented by \citet{nkk00} and \citet{bif01}.

In the context of these models, one may envisage the accretion disk of H~1821+643 to be comprised of a highly ionized optically thin atmosphere sitting atop a mostly neutral disk. The Thomson depth of the upper Compton heated layer should be such (e.g. $\tau_{T} \lesssim 1$) that most of the X-rays propagating through the disk can reach the bottom colder layers to produce the observed cold reflection spectrum, while also showing ionized features. (\citealp{nkk00} and \citealp{bif01} report narrow transitions from hot to cold material in the  illuminated layers.) It is curious that a He-like Fe line at $\sim 6.7$~keV is not seen. However, it is possible that the upper layer has an ionization state to support only H-like Fe. 

\section{Absorption \label{sec:abs}}

\subsection{Constraints on Absorption Column Densities \label{sec:constraint}}

H~1821+643 is one of the brightest X-ray quasars at low redshift, therefore it provides a unique opportunity to probe the so-called ``missing baryons'' along this particular line-of-sight (see \citealp{fhp98} and \citealp{cos99a} for discussion about the missing baryons problem). The path length to H1821+643 is almost $1h^{-1}$ Gpc. The low Galactic hydrogen column density ($\sim 4\times 10^{20}\ {\rm cm^{-2}}$) and relatively simple spectral shape reduce the spectral complexity typically seen in Seyfert galaxies and other active galaxy nuclei, making it easy to identify any intervening absorption. Furthermore, recent observations with {\sl HST} and {\sl FUSE} indicate that there are at least six intervening \ion{O}{6} absorption systems (\citealp{ots00}; see Table~\ref{tbl-3}). Highly ionized ions, such as \ion{O}{6}, are important ion species in diagnosing the physical conditions of the intergalactic medium (IGM). The production of \ion{O}{6} ions requires photons or electrons with energy $> 114$ eV, so \ion{O}{6} can be produced by photoionization from the background radiation, or collision ionization in a hot gas with temperatures above $10^{5}$ K. However, current observations of \ion{O}{6} absorption lines cannot definitively constrain the ionization mechanism \citep{tsa00}. X-ray absorption lines from He-like or H-like ions, such as \ion{O}{7} (0.574 keV) and \ion{O}{8} (0.654 keV) can give direct evidence of a hot, collisionally ionized gas with temperature ranging from $5\times 10^{5}$ K to $10^{7}$ K.

\placetable{tbl-3}

Identifying absorption features requires a careful measurement of the continuum. To achieve this, we take the MEG data (HEG does not have enough effective area below 0.8 keV) and initially fit the 0.4---4 keV spectrum with a power law and the Galactic absorption. We then fit the residual with a five-order polynomial to obtain an accurate characterization of the continuum. The five-order polynomial will account for spectral features larger than $5\AA$ (such as calibration uncertainties) but will preserve narrow line features. The data are binned at $0.02\AA$ and we calculated the $\chi$ (the signal-to-noise ratio) of each bin. Figure~\ref{fig-5} shows the $16 - 28 \AA$ spectrum. The bottom panel of each plot gives $\chi$, and the two dotted lines within the $\chi$ plot correspond to $\pm3\sigma$ level.

\placefigure{fig-5}

An absorption line feature should at least have a signal-to-noise ratio of $< -3\sigma$ level to be identified. Figure~\ref{fig-5} shows that each bin in 16---28$\AA$ has $\chi > -3$, which indicates that no absorption line is identified in this observation. Interestingly, we find an emission feature at around $21 \AA$ with $\chi \sim 3.5$. There is no known emission feature in this wavelength in the rest frame of the quasar. Assuming a Poisson distribution for each bin, the probability for observing one bin with $\chi \ge 3.5$ in one observation of H~1821+643 is $\sim 50\%$. So the line may be the result of a statistical fluctuation.

Given the six intervening \ion{O}{6} absorption systems in the UV spectra of H~1821+643 and assuming that X-ray absorption ions coexist with \ion{O}{6}, we calculate the upper limits of \ion{O}{7} and \ion{O}{8} column densities, and place constraints on the physical conditions of the absorption systems. The $3\sigma$ upper limits of the \ion{O}{7} and \ion{O}{8} column densities for all six intervening systems are estimated based on the quasar spectral flux, the MEG resolution and effective area. We also take into account the instrumental line response function (see \citealp{fmb01}). The results are shown in Table~\ref{tbl-3}. We take a constant velocity dispersion of $b = 100\ {\rm km\ s^{-1}}$. Hydrodynamic simulations \citep{fbc01} predict that the the mean \ion{O}{7} and \ion{O}{8} Doppler $b$-parameters are around $\sim 50\ {\rm km\ s^{-1}}$ and the turbulence in the got gas can easily increase $b$-parameter to $\sim 100\ {\rm km\ s^{-1}}$. Since the MEG has very little collection area below 0.4 keV, we estimate the upper limit of \ion{O}{7} column density only with the $z=0.1214$ system. The \ion{O}{6} column densities were adopted from various references (\citealp{tsj00, ots00, stl98}). For those observations which had column densities of the \ion{O}{6} $\lambda\lambda\ 1031.93\AA, 1037.62\AA$ doublet, we adopt the average of the two values.

\subsection{A Sample Study: $z\approx 0.1214$ System \label{sec:0.1214}}

Our limits on \ion{O}{7} and \ion{O}{8} absorption constrains the physical condition of the \ion{O}{6} absorber if it is collisionally ionized, as recent hydrodynamic simulations predict (\citealp{fbr01, cto01}), but not if it is photoionized. We take the $z\approx 0.1214$ absorption system as an example. The \ion{O}{6} absorption ($z = 0.12137$; $EW=98 \pm 21\ {\rm m\AA}$) was detected by \citep{ots00} with {\sl FUSE}. They also detected an \ion{H}{1} Ly$\beta$ absorption line at $z=0.12129$, with an EW of $137 \pm 23\ {\rm m\AA}$ and a column density of $3.2 \pm 1.6\times 10^{14}\ {\rm cm^{-2}}$.

{\bf Collisionally Ionized Absorber: } We find that if the \ion{O}{6} absorption lines are produced in hot gas by collisional ionization, the ratio between \ion{O}{6} and \ion{O}{7} or \ion{O}{8} column densities can place tight constraints on the gas temperature. Figure~\ref{fig-6} displays $\log (N_{OVI}/N_{OVII})$ and $\log (N_{OVI}/N_{OVIII})$. The atomic data was adopted from \citet{mmc98}. The $3\sigma$ upper limits on \ion{O}{7} or \ion{O}{8} column densities are $8.81 \times 10^{15}\ {\rm cm^{-2}}$ and $2.27 \times 10^{16}\ {\rm cm^{-2}}$ respectively, which give a lower limit of $\log (N_{OVI}/N_{OVII}) > -2.27$ and $\log (N_{OVI}/N_{OVIII}) > -2.36$. Figure~\ref{fig-6} shows that the gas temperature must be lower than $10^{6}$ K to satisfy the \ion{O}{6}-\ion{O}{7} ratio and lower than $2\times 10^{6}$ K to satisfy the \ion{O}{6}-\ion{O}{8} ratio. This gives an upper limit of the temperature of the $z\approx 0.1214$ absorption system at $T < 10^{6}$ K. We can safely place a lower limit on the temperature of $T>10^{5}$ K according to the collisional ionization fraction of \ion{O}{6} \citep{mmc98}. 

\placefigure{fig-6}

The temperature range, $10^{5} < T < 10^{6}$ K, is in accord with cosmological simulations that predict most of these \ion{O}{6} absorbers are located in the filaments that connect the virialized regions and have been shock-heated to temperatures about $10^{5}$ K. Recently large-scale cosmological hydrodynamic simulations have shown that a large amount of the intergalactic medium ($30\%\sim50\%$) lies in regions with overdensity $5 < \delta < 200$ and temperature $10^{5} < T < 10^{7}$ K (Warm-hot intergalactic medium, or ``WHIM'', see, e.g., \citealp{cos99a, dco01, fbc01}). Here $\delta = \rho_{b}/\left<\rho_{b}\right> - 1$ and $\left<\rho_{b}\right>$ is the mean baryon density of the universe. In this temperature range, most of the gas should be collisionally ionized. \citet{fbr01} and \citep{cto01} showed that by assuming collisional ionization in WHIM gas, hydrodynamic simulation can reasonably reproduce the observed distribution of \ion{O}{6} absorption lines. Furthermore, \citet{fbr01} and \citet{cto01} showed that collisional ionization dominates for \ion{O}{6} absorption lines with $EW > 35-40\ {\rm m\AA}$. 

{\bf Photoionized Absorber:} If the \ion{O}{6} is produced by photoionization in a low density gas, corresponding column densities of \ion{O}{7} and \ion{O}{8} are too low to be detected with any current X-ray telescope. We used CLOUDY \citep{fer01} to calculate the ionization structure in a slab of gas illuminated by a background radiation from \citet{hma96}, in which we adopt a mean specific intensity at the Lyman limit of $J_{\nu} = 2\times10^{-23}\ {\rm ergs\ s^{-1}Hz^{-1}sr^{-1}}$. Three metallicity models are considered here: two with a constant metallicity ($Z=0.1Z_{\odot}$ and $1Z_{\odot}$, where $Z_{\odot}$ is the solar abundance) and one in which metallicity is a function of baryon density. In the last model, we adopt the numerical simulation data from \citet{cos99b}. The CLOUDY calculation stops when the neutral hydrogen column density reaches $\log N_{HI} = 14.51$ \citep{ots00}. To obtain the \ion{O}{6}, \ion{O}{7} and \ion{O}{8} column densities under different physical conditions we adjust the ionization parameter $U=n_{\gamma}/n_{H}$, where $n_{\gamma}$ is the density of \ion{H}{1} ionizing photons and $n_{H}$ is the hydrogen number density. 

We find that none of the three models can give detectable \ion{O}{7} and \ion{O}{8} column densities in the X-ray band. The results are shown in Figure~\ref{fig-7}. The top, middle and bottom panels stand for models with $Z=0.1Z_{\odot}$, $Z=1Z_{\odot}$ and $Z=Z(\rho_{b})$ ($\rho_{b}$ is the baryon density, see \citealp{cos99b}), respectively. For each model we calculate the \ion{O}{6}, \ion{O}{7} and \ion{O}{8} column densities. The observed \ion{O}{6} column density is shown as the horizontal dashed line in each panel and the vertical dashed line in each panel indicates the corresponding ionization parameters (or hydrogen number density). These vertical lines also give corresponding \ion{O}{7} and \ion{O}{8} column densities. None of these \ion{O}{7} and \ion{O}{8} column densities can be detected with current and planned X-ray telescopes (for X-ray telescope detection limits, see \citealp{fca00}). For instance, in the $Z=0.1Z_{\odot}$ model, the expected \ion{O}{7} and \ion{O}{8} column densities are only $\sim 10^{14}\ {\rm cm^{-2}}$ and $< 3.2 \times 10^{12}\ {\rm cm^{-2}}$, respectively. Given such a low density system at $z\approx 0.1214$, only when its metallicity reaches solar abundance could \ion{O}{7} absorption line be detected with current X-ray instruments (When $-5.4 < \log(n_{H}) < -6.4$; see Figure~\ref{fig-7}). Recent hydrodynamic simulations showed that the typical metallicity of the gas that produces \ion{O}{6} absorption lines is $0.1-0.3\ Z_{\odot}$ \citep{cto01}. So we conclude that any X-ray absorption lines that are discovered by current telescopes are unlikely to come from a low-density gas photoionized by the background radiation.

\placefigure{fig-7}

\section{The Surrounding Cluster \label{sec:cluster}}

Optical studies revealed that H~1821+643 lies in a rich cluster of galaxies (\citealp{sbg92, lrh92}). Optical spectroscopy of six member galaxies showed that the redshift of the cluster, $0.299\pm0.002$, is consistent with the redshift of the quasar \citep{sbg92}. They also determined the velocity dispersion of the cluster to be $1050\pm320\ {\rm km\ s^{-1}}$ which indicates that this cluster is quite massive. This is consistent with the cluster richness, R$>$2, determined by \citet{lrh92}.

Figure~\ref{fig-1} shows the zeroth order image, and from the contours it appears that there is extended emission around H~1821+643. To quantify this we first corrected the image for the known detector irregularities and for telescope vignetting. We then extracted a profile in radial bins with each bin being $1^{\arcsec}$ wide.  The profile from 2$^{\arcsec}$ to 160$^{\arcsec}$ is shown in Figure~\ref{fig-8} along with the model fits. The initial model was a combination of the Chandra point spread function (PSF) and a $\beta$-model \citep{cff78} for the extended emission from the cluster. This proved to be a rather poor fit to the X-ray profile, leaving excess emission between 5 and 15$^{\arcsec}$. We then added a Gaussian component to account for this extra emission and found an acceptable fit to the profile. Formally, the reduced $\chi^2$ is 2.51 (343.6/137 dof) but the largest contribution is from the quasar where a modest amount of instrumental pileup is not well modeled by the Chandra PSF. Evaluating the model between 5$^{\arcsec}$ and 160$^{\arcsec}$ results in $\chi^2_\nu$ = 1.71. 

\placefigure{fig-8}

The parameters for the cluster profile are similar to those found for other clusters, a core radius of $17.6^{+0.17}_{-0.17}\ ^{\arcsec}$ (0.1 Mpc) and  a $\beta = 0.74^{+0.05}_{-0.03}$. The additional Gaussian component has a $\sigma$ value of $6.54^{+0.14}_{-0.13}\ ^{\arcsec}$. At the distance of the cluster this corresponds to 37.3 kpc, which is similar to the scale for a giant elliptical galaxy. 

The ACIS-S3 spectral data for the cluster was extracted from $3^{\arcsec}$ to $100^{\arcsec}$ and this is what we refer to as the global spectrum (Figure~\ref{fig-9}). By excluding a 3$^{\arcsec}$ radius circle around the quasar we have eliminated essentially all the flux from a point source from our analysis (the {\sl Chandra} Proposers' Observatory Guide, Figure 6.3). We used the XSPEC {\sc mekal} model with Galactic absoption to fit the cluster spectrum and the fit parameters are given in Table~\ref{tbl-4}. In addition to the temperature and abundance we also allowed the redshift to be a free parameter in fitting the global spectrum. From the X-ray data we find that the redshift of the hot gas is 0.303 with a lower limit of 0.299 and an upper limit of 0.309 ($90\%$ confidence limits) consistent with the redshift of the quasar and that of the optical galaxies. We also had sufficient counts to extract the spectrum in radial rings and the parameters for these fits are listed in Table~\ref{tbl-4}. We find the overall temperature and metal abundance ($kT = 10.8^{+1.0}_{-0.9}$ keV and $Z = 0.35\pm0.08\ Z_{\odot}$) are consistent with those of a typical hot cluster. The luminosity between 2 and 10 keV (quasar frame) is $2.54 \times 10^{45}\ {\rm ergs\ s^{-1}}$, which is roughly consistent with temperature-luminosity relationship obtained with {\sl ASCA} \citep{hbg00}. We also find that the temperature is cooler in inner regions between 3 and $11.5^{\arcsec}$; however, the cooler temperature is not consistent with that of a typical elliptical galaxy \citep{dwh96}.

\placetable{tbl-4}
\placefigure{fig-9}

We estimate that the cluster makes negligible contribution to the iron 6.9 keV emission line detected in the grating spectrum. The strongest emission line from a hot cluster plasma in that line region is \ion{Fe}{26} K$\alpha$ line. From the 3 to $11^{\arcsec}.5$ region, we obtain the \ion{Fe}{26} K$\alpha$ line flux is $\sim 5.2 \times 10^{-7}\ {\rm photons\ cm^{-2}s^{-1}}$. We also estimate that for the part of the cluster hidden by the quasar the fraction of the cluster flux should be at most $6.56\%$ from the surface brightness profile (Figure~\ref{fig-8}). By comparing the line flux from the grating spectra in Table~\ref{tbl-2}, we find that the cluster can contribute at most $\sim 3\%$ to the 6.9 keV line flux.

\section{Summary \label{sec:summary}}

In this paper we study the spectral features of a low-redshift quasar H~1821+643 with {\sl Chandra} HETGS. Our main conclusions can be summarized as follows:

\begin{enumerate}

\item An emission feature at $\sim$ 5 kev was detected at nearly $6\sigma$ level and has been identified as a fluorescent iron line. The rest-frame line-center energy ($6.435 \pm 0.041$ keV) implies this line is from neutral or low ionization states of iron, although we cannot rule out contribution from ionization states up to \ion{Fe}{19}. An emission feature at around 6.9 keV (rest frame of the quasar) was marginally detected at $2.3\sigma$ level and is identified as a possible H-like Fe emission line.

\item The {\sc diskline} model provides an acceptable fit to the 6.4 keV iron line; however, a putative torus could also make contributions to the EW of the emission line. The model in which X-rays are reflected from the BLR clouds requires the X-ray flux from H~1821+643 to vary at a significant level (a factor of at least 5) on a timescale of $\sim 2$ months, which seems unlikely. In order to accommodate both the Fe~K$\alpha$ and \ion{Fe}{26} lines, we suggest that both lines could originate in an accretion disk comprised of a highly ionized optically thin atmosphere sitting atop a mostly neutral disk.

\item No absorption features were detected at or above the $3\sigma$ level. A total of six \ion{O}{6} intervening absorption systems have been detected with {\sl HST} and {\sl FUSE}. We place $3\sigma$ upper limits on \ion{O}{7} and \ion{O}{8} column densities at the corresponding redshifts, which have typical values of $\sim 10^{16}\ {\rm cm^{-2}}$.

\item We focus on the $z\approx 0.1214$ absorption system and constrain its physical conditions by combining UV and X-ray observations. Assuming collisional ionization, we constrain the gas temperature at $10^{5} < T < 10^{6}$ K, which is consistent the results from hydrodynamic simulations. However, if the \ion{O}{6} absorbers are photo-ionized by the background radiation, no detectable \ion{O}{7} or \ion{O}{8} lines are expected. 

\item We obtain the surface brightness profile and spectra of the cluster that surrounds H~1821+643. The X-ray spectra reveal that this is a typical hot cluster with a temperature of $kT \sim 10$ keV and a metal abundance of $\sim 0.3Z_{\odot}$. We also obtain the redshift of the cluster, which is consistent with results from optical measurements. We estimate that the cluster makes negligible contributions to the 6.9 keV iron K line flux. An additional Gaussian component is found in fitting the surface brightness profile. This component has a width of that similar to the scale for a giant elliptical galaxy.

\end{enumerate}

\acknowledgments{TF thanks the MIT/CXC team for its support. This work is supported in part by contracts NAS 8-38249 and SAO SV1-61010. Support for GLB was provided by NASA through Hubble Fellowship grant HF-01104.01-98A from the Space Telescope Science Institute, which is operated by the Association of Universities for Research in Astronomy, Inc., under NASA contract NAS 6-26555.}


\clearpage
\figcaption[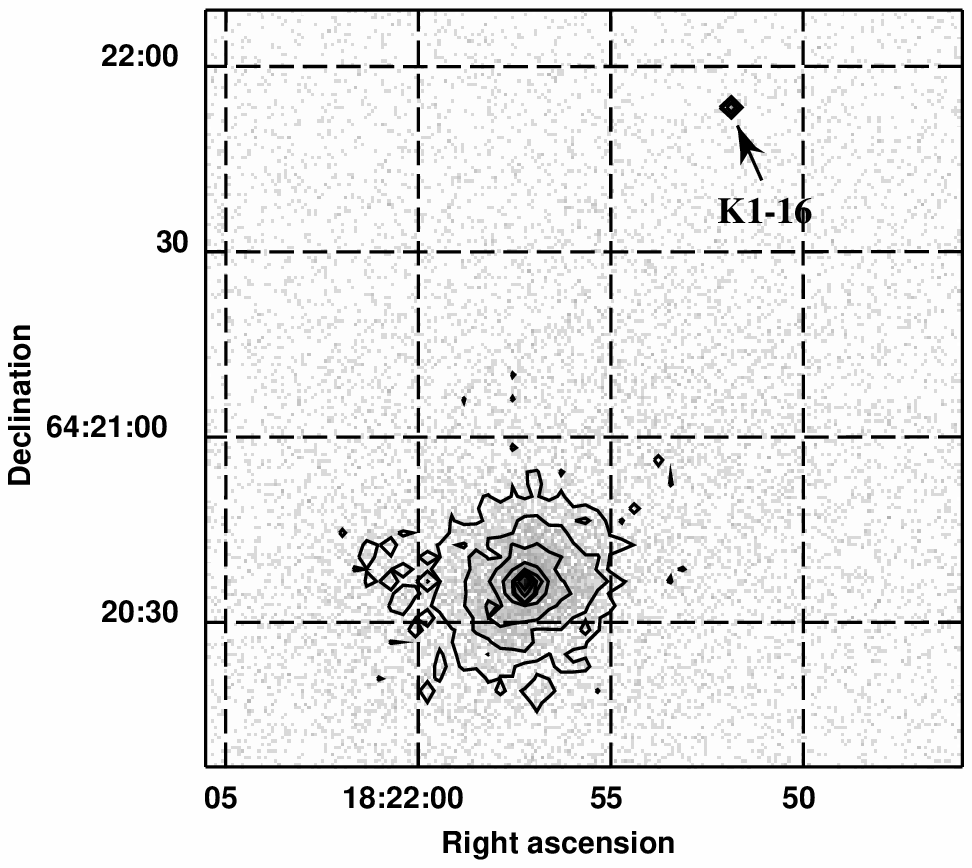]{the zeroth-order image of H~1821+643 in a $2^{\arcmin} \times 2^{\arcmin}$ region, with North to the top and East to the left. The image has been smoothed by evaluating the contour at every $3$ image pixels for visual clarity. The X-ray contours are separated by a factor of $\sim 2.8$ in surface brightness, with the lowest contour at a level of $\sim 0.7\ {\rm photons\ arcsec^{-2}}$. The star K1-16 is indicated. From the contours it appears that there is extended emission around the central source.\label{fig-1}}
\figcaption[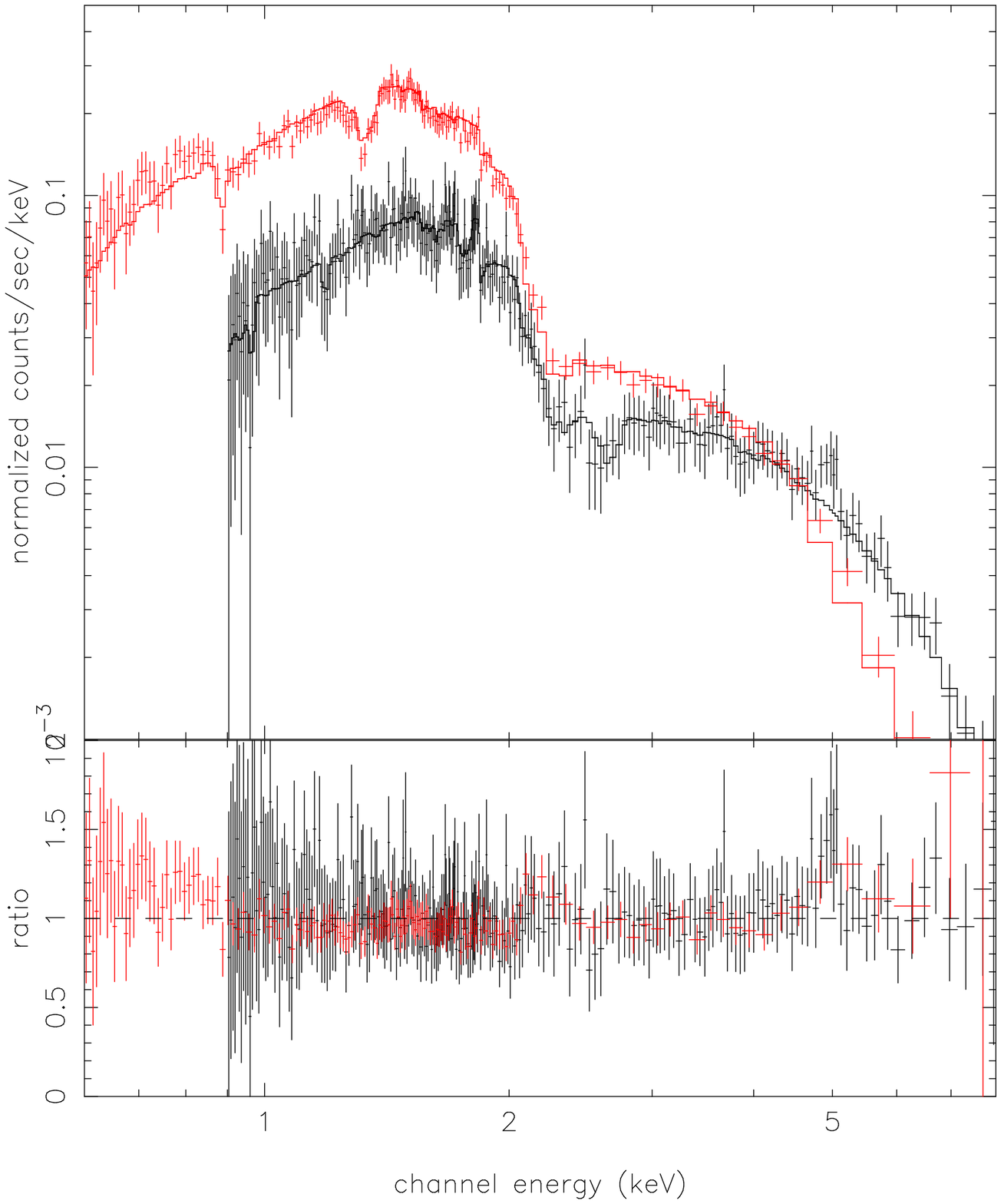]{The first order HETGS spectra of H~1821+643 (red: MEG spectrum; black: HEG spectrum). Both spectra are fitted simultaneously with a single power law plus photoelectric absorption at $z=0$ (see Table~\ref{tbl-1} for fitting parameters). An emission line feature is clearly identified around 5 keV. Residuals of a weak emission line are also present around 5.3 keV. \label{fig-2}}
\figcaption[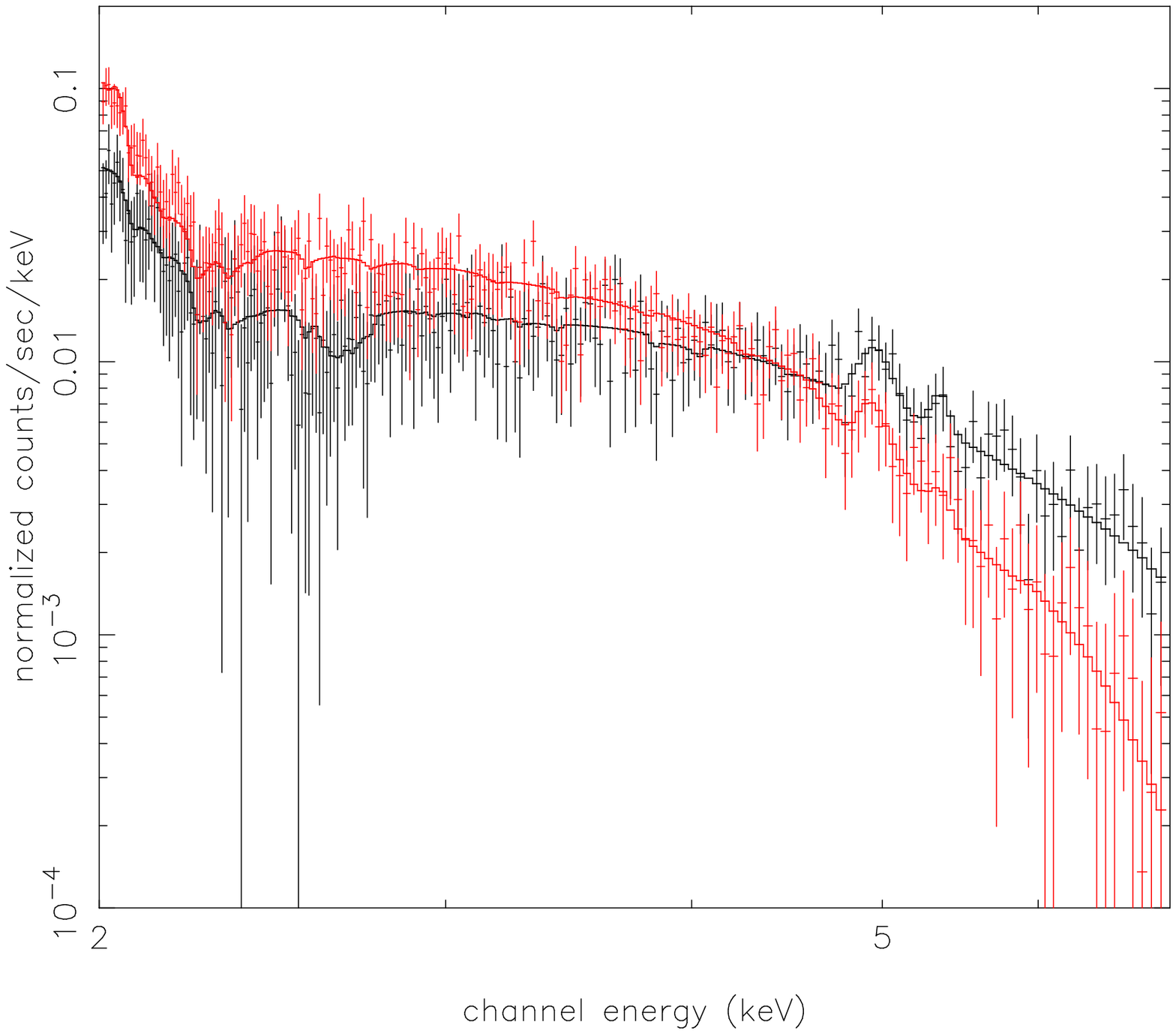]{The simultaneous fit to MEG (red) and HEG (black) spectra between 2 and 7 keV. Two Gaussian components are added to fit the emission features around 5 and 5.3 keV (see Table~\ref{tbl-2} for fitting parameters). \label{fig-3}}
\figcaption[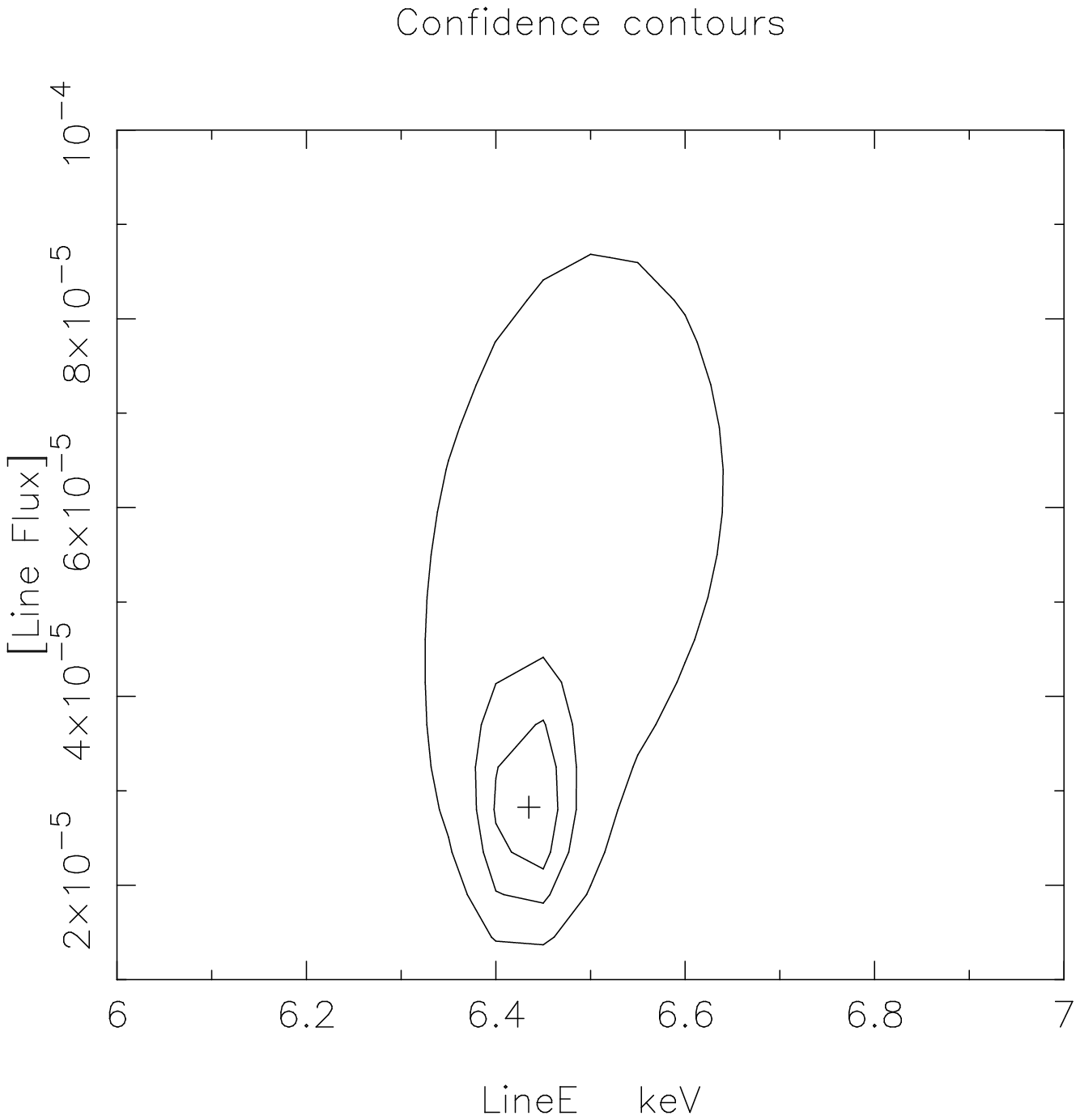]{The $68\%$, $90\%$ and $99\%$ joint confidence regions (for two parameters of interest) for the line center energy (rest frame) and the line flux. The central cross indicates for the best-fit values for both parameters. \label{fig-4}}
\figcaption[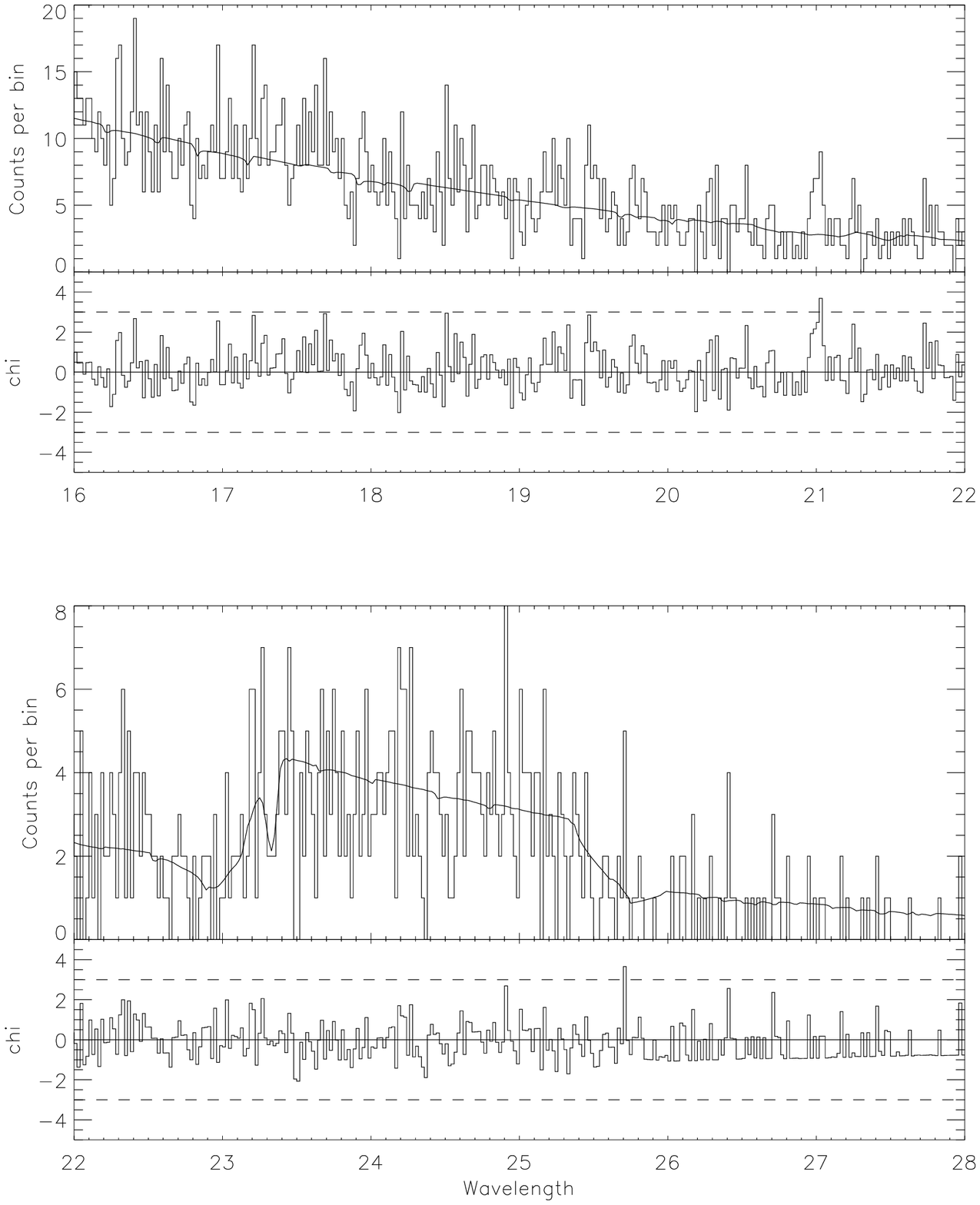]{The first order MEG spectrum between 16 and 28 $\AA$. The binsize is 0.02 $\AA$. The spectrum is initially fitted with a power law and the Galactic absorption. The residual is then fitted with a five-order polynomial to obtain an accurate characterization of the continuum. The bottom plot in each panel gives the $\chi$ of each bin, and the dashed lines indicates $\pm 3\sigma$ level. \label{fig-5}}
\figcaption[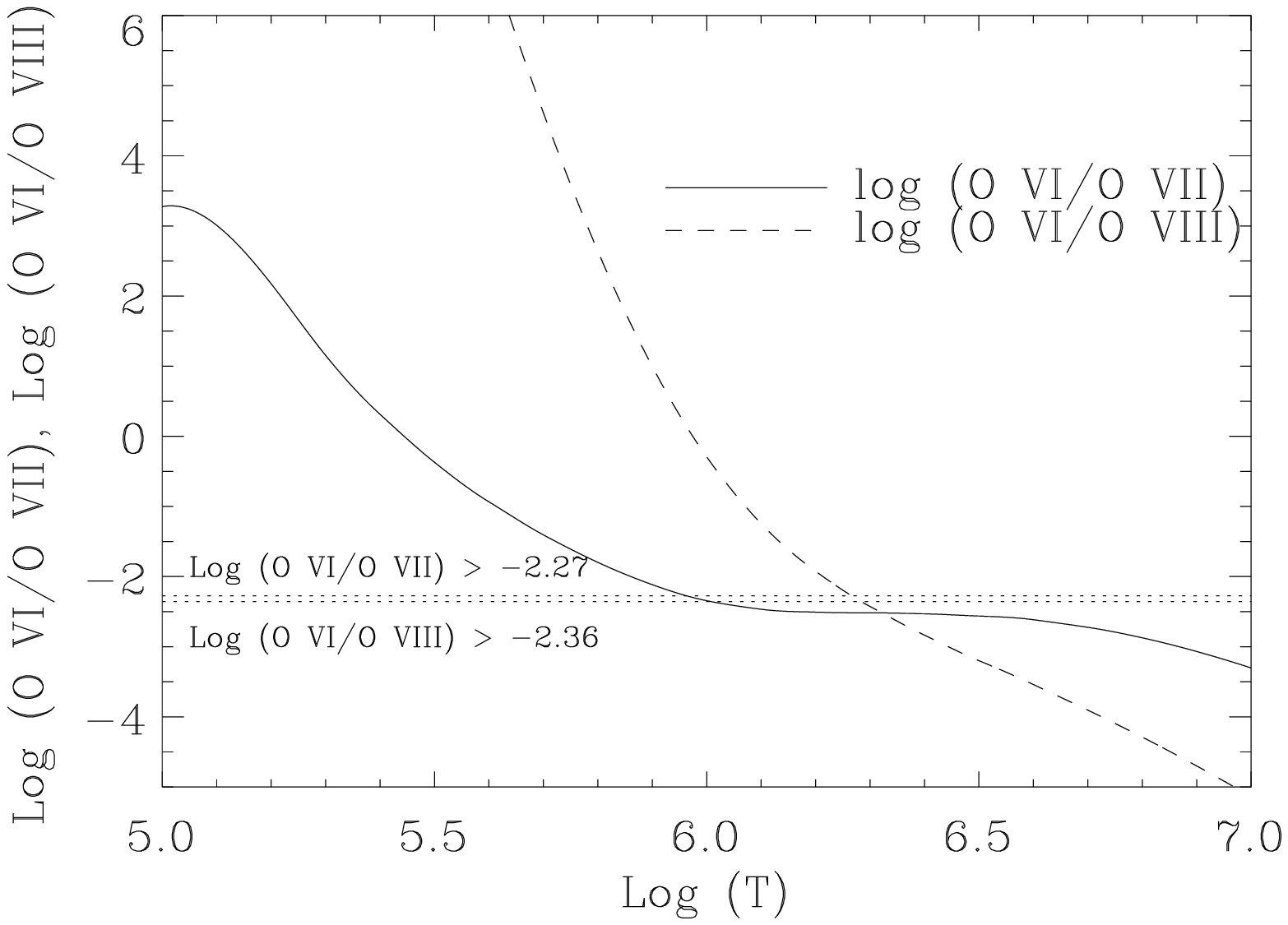]{Collisional Ionization Model: the O VI/O VII (solid line) and O VI/O VIII (dashed line) ratio vs. temperature in the $z\approx 0.1214$ system. The horizontal lines indicate the observed lower limit on O VI/O VII and O VI/O VIII, and give the corresponded upper limits on temperatures. \label{fig-6}}
\figcaption[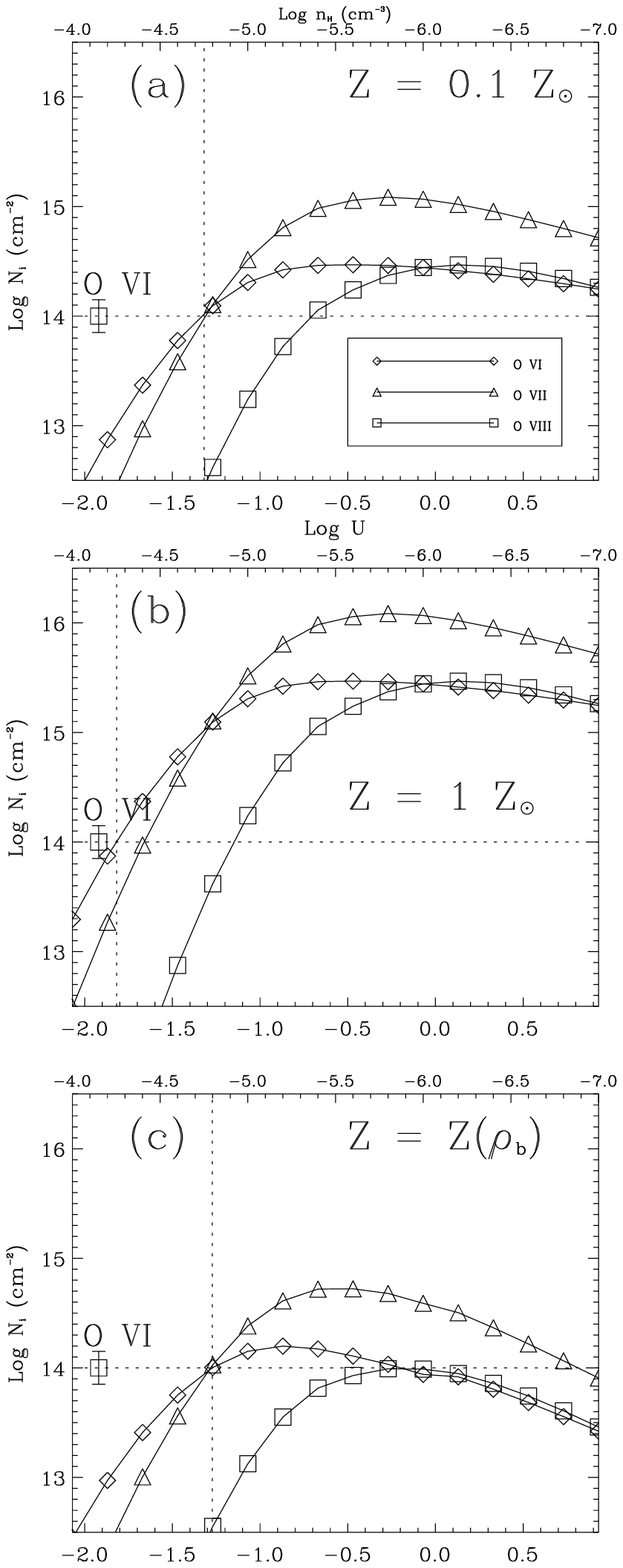]{Photoionization Model: the O VI, O VII and O VIII column densities vs. the ionization parameter in the $z\approx 0.1214$ system. The corresponded hydrogen number density is also labeled. The three panels corresponds to three metal abundances (a) $Z=0.1Z_{\odot}$, (b) $Z=1Z_{\odot}$ and (c) $Z=Z(\rho_{b})$, where $Z(\rho_{b})$ is from \citet{cos99b}. The horizontal lines in each panel indicate the observed O VI column density. From the three vertical lines one can find the corresponding O VII and O VIII column densities in the system which produces the observed O VI column density. None of the three models can give detectable O VII and O VIII column densities in the X-ray band with the current instruments. \label{fig-7}}
\figcaption[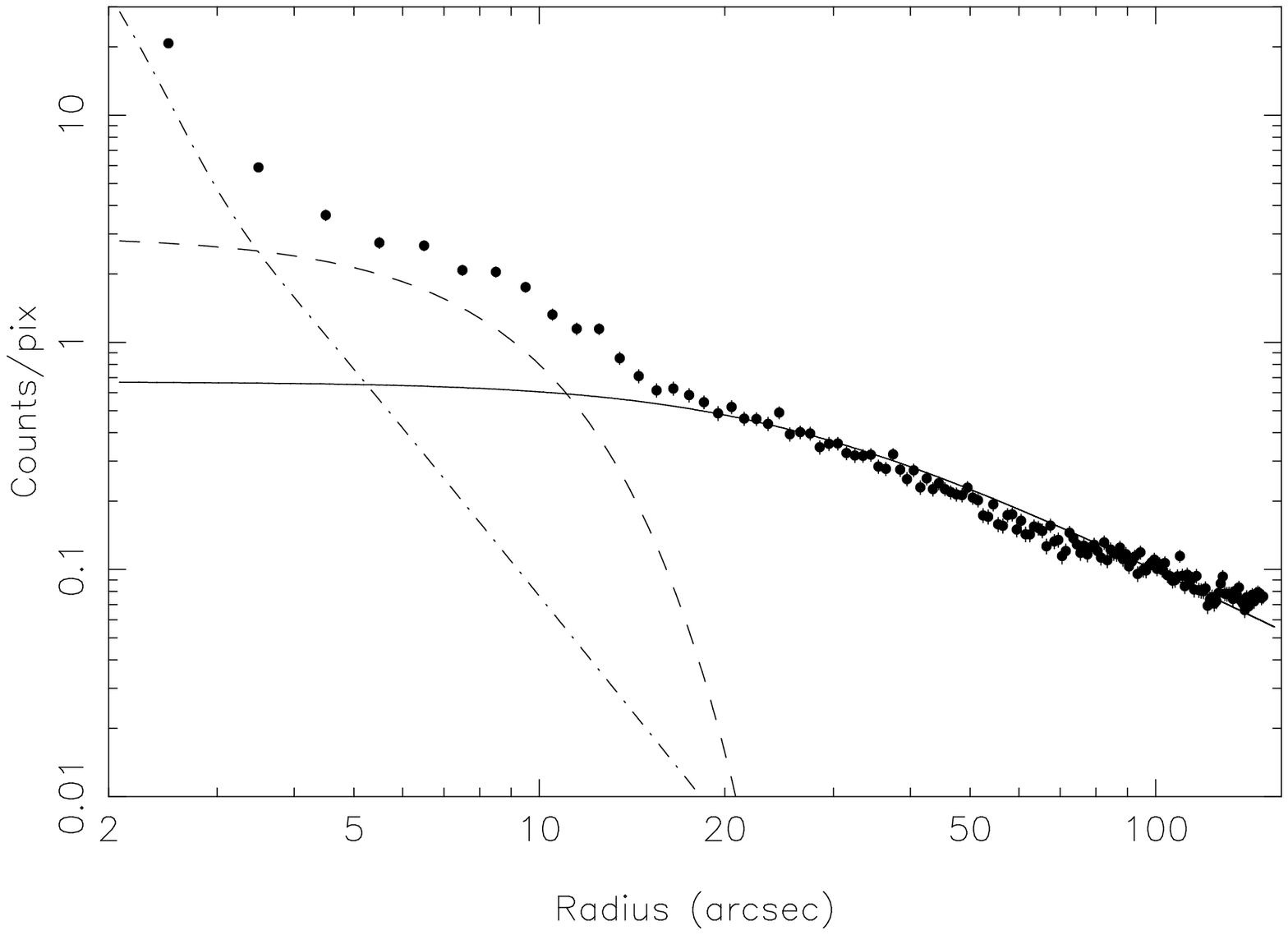]{The surface brightness profile of H~1821+643 and the surrounding cluster. The data (solid dots) is fitted with a model containing three components: the {\sl Chandra} PSF in the center to account for the point source (dot-dashed line), a $\beta$-model to account for the extended cluster emission (solid line) , and a Gaussian component with a $\sigma \sim 6.45\arcsec$ (dashed line). Data from less then $2^{\arcsec}$ are affected by pileup and are not included \label{fig-8}}
\figcaption[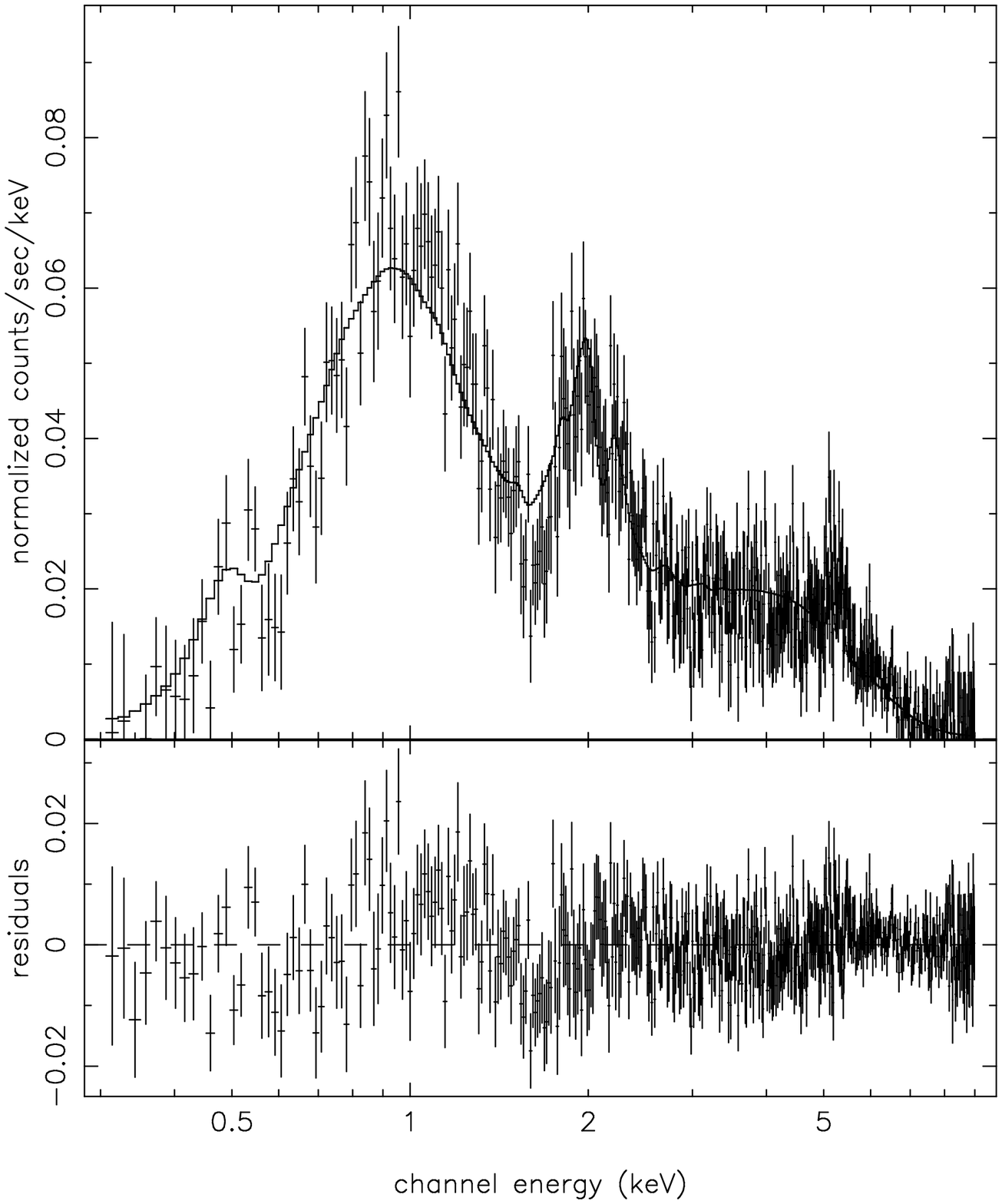]{The zeroth order spectrum and the residual (extracted from the region with an inner radius of $3\arcsec$ and an outer radius of $100\arcsec$. the XSPEC {\sc mekal} model with Galactic absorption is adopted to fit the cluster spectrum and the fit parameters are given in Table~\ref{tbl-4}. \label{fig-9}}


\clearpage
\begin{deluxetable}{lccccrc}
\tabletypesize{\footnotesize} 
\tablewidth{0pt}
\tablecaption{Continuum Spectral Fit \label{tbl-1}}
\tablehead{\colhead{Energy Range} & \colhead{$N_{H}$} & \colhead{Photon Index} &
\colhead{$A_{pl}$\tablenotemark{a}} & \colhead{C-statistic/dof} & 
\colhead{$f_{2-10}$\tablenotemark{b}} & \colhead{$\log L$\tablenotemark{c}} \\
\colhead{(keV)} & \colhead{($10^{20}\ {\rm cm^{-2}}$)} & \colhead{($\Gamma$)} & & & & }
\startdata 
0.9-8 (HEG)  0.6-8 (MEG) & 3.8 (fixed) & $1.844 \pm 0.017$         & $3.65 \pm 0.04$ & $1906/1567$ & $1.19 \pm 0.01$ & $45.690 \pm 0.002$  \\
                         & $ < 0.19$ & $1.771 \pm 0.017$         & $3.65 \pm 0.04$ & $1824/1566$ & $1.33^{+0.02}_{-0.01}$ & $45.731 \pm 0.004$\\
2-7 (HEG, MEG)           & 3.8 (fixed) & $1.761^{+0.047}_{-0.052}$ & $3.40 \pm 0.20$ & $537/440$   & $1.27^{+0.06}_{-0.09}$ & $45.700 \pm 0.030$ \\
\enddata
\tablenotetext{a}{Flux at 1 keV (observer's frame) in units of $10^{-3}\ {\rm photons\ cm^{-2}s^{-1}\ keV^{-1}}$}
\tablenotetext{b}{Absorbed flux between 2 and 10 keV (observer's frame) in units of $10^{-11}\ {\rm ergs\ cm^{-2}s^{-1}}$}
\tablenotetext{c}{Intrinsic luminosity between 2 and 10 keV (quasar frame) in units of ${\rm ergs\ s^{-1}}$, $q_{0} = 0.5$}
\tablenotetext{d}{All errors are quoted at $90\%$ confidence for one interesting parameter ($\Delta C=2.71$)}
\end{deluxetable}

\clearpage
\begin{deluxetable}{llll}
\tablecaption{Gaussian Line Fit \label{tbl-2}}
\tablehead{\colhead{} & \colhead{Parameter} & \colhead{HEG only} & \colhead{HEG \& MEG}}
\startdata
Gaussian Line Model & Line energy (keV) & $6.445^{+0.052}_{-0.061}$ & $6.435 \pm 0.041$ \\
 & Line width  (eV)  & $114.2^{+63.3}_{-45.4}$   & $106.9^{+58.3}_{-37.8}$ \\
 & Line flux ($10^{-5}\ {\rm photons\ cm^{-2}s^{-1}}$) & $3.03^{+1.24}_{-1.26}$ & $2.88^{+1.07}_{-0.91}$ \\
 & EW (eV) & $116.0^{+47.5}_{-48.2}$ & $113.0^{+42.1}_{-35.7}$ \\
 & C-stat/dof & 244.2/217 & 496.8/437 \\
 & Line energy (keV) & $6.932 \pm 0.061$ & $6.942^{+0.050}_{-0.068}$ \\
 & Line width  (eV)  & ...   & $63.9^{+71.4}_{...}$ \\
 & Line flux ($10^{-6}\ {\rm photons\ cm^{-2}s^{-1}}$) & $4.90^{+7.20}_{-4.40}$ & $1.08^{+0.69}_{-0.70}$ \\
 & EW (eV) & $21.1^{+31.0}_{-18.9}$ & $48.6^{+31.2}_{-31.4}$ \\
 & C-stat/dof & 240.9/214 & 487.1/434 \\
\cline{1-4} 
{\sc diskline} Model & Line energy (keV) & $6.49^{+0.10}_{-0.08}$ & $6.49 \pm 0.06$  \\
 & Inclination ($\theta$, degree) & $21.49^{+6.90}_{-10.40}$ & $20.20^{+6.16}_{-7.35}$ \\
 & Line flux ($10^{-5}\ {\rm photons\ cm^{-2}s^{-1}}$) & $3.24^{+1.14}_{-1.38}$ & $3.04^{+0.94}_{-0.89}$\\
 & EW (eV) & $168.0^{+59.1}_{-71.6}$ & $161.0^{+49.8}_{-47.1}$ \\
 & C-stat/dof & 243.1/217 & 497.1/437 \\

\enddata
\end{deluxetable}

\clearpage
\begin{deluxetable}{llllc}
\tablewidth{0pt}
\tablecaption{Oxygen Column density\tablenotemark{a} \label{tbl-3}}
\tablehead{\colhead{Redshift} & \colhead{\ion{O}{6}} & \colhead{O VII\tablenotemark{b}} & \colhead{O VIII\tablenotemark{b}} & \colhead{Reference\tablenotemark{c}}}
\startdata
0.1214 & 14.00 & $<15.94$ & $<16.36$ & [1] \\
0.2133 & 13.55 & ...   & $<16.37$ & [2] \\
0.2249 & 14.31 & ...   & $<16.37$ & [3] \\ 
0.2267 & 13.48 & ...   & $<16.26$ & [2] \\
0.2453 & 13.79 & ...   & $<16.51$ & [2] \\
0.2666 & 13.74 & ...   & $<16.40$ & [2] \\
\tablenotetext{a}{Log column density, in unit of ${\rm cm^{-2}}$.}
\tablenotetext{b}{The $3\sigma$ upper limits of the \ion{O}{7} and \ion{O}{8} column densities.}
\tablenotetext{c}{[1]: \citet{ots00}; [2]: \citet{tsj00}; [3]: \citet{stl98}}
\enddata
\end{deluxetable}

\clearpage
\begin{deluxetable}{lcclr}
\tabletypesize{\footnotesize} 
\tablewidth{0pt}
\tablecaption{Cluster Fits \label{tbl-4}}
\tablehead{\colhead{Region} & \colhead{Temperature} & \colhead{Abundance} & \colhead{$\chi^{2}$/dof} \\
\colhead{} & \colhead{(keV)} & \colhead{(Solar)} & \colhead{}}
\startdata
3. -- 100$\arcsec$ &10.8$^{+1.0}_{-0.9}$&0.35$^{+0.08}_{-0.08}$&571.3/488\\
3. -- 11.5$\arcsec$ &4.5$^{+0.5}_{-0.4}$&0.36$^{+0.10}_{-0.10}$&196.2/107\\
17.5 --50.$\arcsec$ &11.3$^{+2.2}_{-1.5}$&0.29$^{+0.11}_{-0.10}$&316.1/267\\
50. -- 100.$\arcsec$&18.7$^{+10.6}_{-4.5}$&0.56$^{+0.33}_{-0.27}$&338.2/323\\
\enddata
\end{deluxetable}

\clearpage
\begin{figure}
\epsscale{1.0}
\plotone{f1.eps}
\end{figure}

\clearpage
\begin{figure}
\epsscale{1.0}
\plotone{f2.eps}
\end{figure}

\clearpage
\begin{figure}
\epsscale{1.0}
\plotone{f3.eps}
\end{figure}

\clearpage
\begin{figure}
\epsscale{0.7}
\plotone{f4.eps}
\end{figure}

\clearpage
\begin{figure}
\epsscale{0.8}
\plotone{f5.eps}
\end{figure}

\clearpage
\begin{figure}
\epsscale{0.9}
\plotone{f6.eps}
\end{figure}

\clearpage
\begin{figure}
\epsscale{0.5}
\plotone{f7.eps}
\end{figure}

\clearpage
\begin{figure}
\epsscale{1.0}
\plotone{f8.eps}
\end{figure}

\clearpage
\begin{figure}
\epsscale{0.7}
\plotone{f9.eps}
\end{figure}

\end{document}